\documentclass[%
 reprint,
 jmp,
superscriptaddress,
 amsmath,amssymb,
 aps,
]{revtex4-2}

\usepackage{graphicx}
\usepackage{dcolumn}
\usepackage{bm}
\usepackage{svg}
\usepackage{braket}
\usepackage{amssymb}

\usepackage{xcolor}
\begin{document}

\preprint{APS/123-QED}

\title{Frequency-bin entanglement-based Quantum Key Distribution}
\author{Noemi Tagliavacche}
\affiliation{Dipartimento di Fisica, Università di Pavia, Via Bassi 6, 27100 Pavia, Italy.}
\author{Massimo Borghi}
\affiliation{Dipartimento di Fisica, Università di Pavia, Via Bassi 6, 27100 Pavia, Italy.}
\email{corresponding author: massimo.borghi@unipv.it} 
\author{Giulia Guarda}
\affiliation{European Laboratory for Non-Linear Spectroscopy (LENS), 50019 Sesto Fiorentino, Italy}
\affiliation{CNR-National Institute of Optics (CNR-INO), 50125 Florence, Italy}
\author{Domenico Ribezzo}
\affiliation{CNR-National Institute of Optics (CNR-INO), 50125 Florence, Italy}
\affiliation{University of L’ Aquila, Department of Physical and Chemical Sciences, 67100 L’ Aquila, Italy}
\author{Marco Liscidini}
\affiliation{Dipartimento di Fisica, Università di Pavia, Via Bassi 6, 27100 Pavia, Italy.}

\author{Davide Bacco}
\affiliation{QTI S.r.l., 50125 Firenze, Italy}
\affiliation{University of Florence, Department of Physics and Astronomy, 50019 Sesto Fiorentino, Italy}

\author{Matteo Galli}
\affiliation{Dipartimento di Fisica, Università di Pavia, Via Bassi 6, 27100 Pavia, Italy.}

\author{Daniele Bajoni}
\affiliation{Dipartimento di Ingegneria Industriale e dell'Informazione, Università di Pavia, Via Ferrata 5, 27100 Pavia, Italy.}

\date{\today}

\begin{abstract}
Entanglement is an essential ingredient in many quantum communication protocols. In particular, entanglement can be exploited in quantum key distribution (QKD) to generate two correlated random bit strings whose randomness is guaranteed by the nonlocal property of quantum mechanics.
Most of QKD protocols tested to date rely on polarization and/or time-bin encoding. Despite compatibility with existing fiber-optic infrastructure and ease of manipulation with standard components, frequency-bin QKD have not yet been fully explored.\\ 
Here we report the first demonstration of entanglement-based QKD using frequency-bin encoding. We implement the BBM92 protocol using photon pairs generated by two independent, high-finesse, ring resonators on a silicon photonic chip. We perform a passive basis selection scheme and simultaneously record sixteen projective measurements. A key finding is that frequency-bin encoding is sensitive to the random phase noise induced by thermal fluctuations of the environment. To correct for this effect, we developed a real-time adaptive phase rotation of the measurement basis, achieving stable transmission over a $26$ km fiber spool with a secure key rate  $\ge4.5$ bit/s. \\
Our work introduces a new degree of freedom for the realization of entangled based QKD protocols in telecom networks.

\end{abstract}

\maketitle


\section{Introduction}
\label{sec:introduction}
The generation and distribution of entanglement at large distances is pivotal to the development of large-scale architectures for distributed quantum computing, metrology and communication \cite{azuma2023quantum}. One of the most successful implementations is quantum key distribution, which enables two remote parties to establish a shared string of bits (key) with unconditional security \cite{xu2020secure}. The security of entanglement-based QKD  comes from the non-local correlations between entangled particles and is guaranteed by the Bell's theorem \cite{nadlinger2022experimental}. Under these assumptions, a secret key can be exchanged between two users even if their  hardware is partially or completely untrusted \cite{zapatero2023advances,pirandola2020advances}. Additionally, employing entanglement relaxes the bounds in security proofs \cite{scarani2009security, pirandola2020advances}. 
In recent years, much progress has been made to close loopholes \cite{nadlinger2022experimental,santagiustina2024experimental}, increase the data rate \cite{ecker2021strategies,mueller2024high}, the communication distance \cite{yin2020entanglement,honjo2008long} and the network size \cite{joshi2020trusted,liu202240,huangsixteen}. Most of the experiments use entangled photon pairs generated by spontaneous parametric down conversion (SPDC) in nonlinear crystals because they have high rates, broadband emission, and great versatility \cite{yin2017satellite,joshi2020trusted,fitzke2022scalable}. Depending on crystal engineering and pump configuration, entangled photon pairs can be created in different degrees of freedom, of which the most exploited are energy-time \cite{joshi2020trusted}, time-bin \cite{fitzke2022scalable} and polarization \cite{shi2020stable}. Their manipulation is easily implemented with table-top solutions such as waveplates and beamsplitters, which can be made cost-effective at telecommunication wavelengths using the widely deployed fiber-optic components.\\ 
In parallel, chip-based quantum key distribution has emerged as a promising direction for making QKD more practical and scalable \cite{liu2022advances}. By integrating most of the components onto a compact photonic chip, chip-based QKD allows for the miniaturization of quantum communication systems, significantly improving their stability and speed of operation, reducing their cost and energy consumption. To date, most of the work focused on the realization of transceiver modules for polarization or time bin encoding using attenuated laser pulses \cite{sibson2017chip,bunandar2018metropolitan}. However, the ever-growing demonstrations of integrated photon pair sources of ultra-high brightness \cite{steiner2021ultrabright}, wide bandwidth of emission \cite{wen2022realizing}, and the generation of entangled photons over multiple degrees of freedom \cite{francesconi2023chip} and in multiple dimensions \cite{zheng2023multichip}, motivate the integration of on—chip entanglement-based QKD solutions. 
For example, recent experiments have demonstrated the transmission of secure keys by using photon pairs generated by SPDC in an AlGaAs nanowire using polarization encoding \cite{appas2021flexible}, or by spontaneous four-wave mixing (SFWM) in an AlGaAs resonator by exploiting energy-time entanglement \cite{steiner2023continuous}. A wide-bandwidth photon pair source from a silicon nanowire has been used in a fully connected network of $40$ users, realized by a combination of beamsplitters and wavelength multiplexing \cite{liu202240}. Similarly, a silicon nitride resonator was used as the central resource of a four-user quantum network \cite{wen2022realizing}. Chip-to-chip entanglement distribution has been demonstrated using path-encoding \cite{llewellyn2020chip} and path-to-polarization conversion in a optical fiber \cite{wang2016chip}. Entanglement has been also distributed by using time-bin qudits, harnessing the intrinsic phase stability of on-chip interferometers in silica \cite{sciara2024chip} and thin-film lithium niobate \cite{mueller2024high}. \\
Much less explored is frequency-bin encoding, i.e. the possibility to encode and manipulate qubits in the frequency degree of freedom \cite{lukens2016frequency}. This strategy has great potential because it is compatible with the existing fiber-optic infrastructure, it is a good candidate for high-dimensional encoding and can be easily manipulated with standard components for telecommunication \cite{lu2023frequency}. However, despite progresses in quantum state engineering \cite{clementi2023programmable,borghi2023reconfigurable}, entanglement distribution and advancements in the implementation of single and multiple qubit gates \cite{lu2020fully}, the use of frequency-bin encoding for QKD has not yet been fully explored. To date, only a feasibility study of entanglement based QKD with frequency bin qubits has been reported, which implements photon pairs generated by a large silicon micro-resonator with a free spectral range of $21$ GHz \cite{henry2024parallelization}. However, the demonstration still has some missing parts, such as random basis selection, simultaneous detection of all the possible measurement outcomes and radio-frequency (RF) clock distribution between the remote users. In addition, the robustness of frequency-bin encoding to both phase noise and bit-flip errors in real-field transmission tests has not been systematically studied.
Here we address these missing points and report the first demonstration of entanglement based QKD using frequency-bin encoding. Entangled photon pairs are generated by two ring resonators of high-finesse on a silicon photonic chip. The use of independent resonators driven by two mutually coherent  pumps \cite{clementi2023programmable,borghi2023reconfigurable} allows us to simultaneously achieve a high pair generation rate of $27\, \textrm{MHz/mW}^2$ and a small frequency-bin spacing of $15$ GHz, suitable for frequency mixing by electro-optic modulation. We implement passive basis selection and simultaneously record all sixteen projective measurements using six superconducting detectors.  A key finding is that the random phase noise induced by thermal fluctuations of the environment impairs transmission after few km of fiber. To correct for this effect, we implement a real-time phase compensation system that keeps the quantum bit error rate (QBER) low and stable over the time.  A systematic investigation of phase noise is performed by correlating the phase drift between frequency-bins, the variation of optical path in the fiber and its temperature. Stable transmission is demonstrated in fiber spools of different lengths up to $26$ km.  
Overall, our work incorporates all the elements to realize frequency-bin QKD with entangled photons in telecom networks. 

\section{Device and experimental setup}
\label{sec:device_and_experimental_setup}
\begin{figure*}[t!]
    \centering
    \includegraphics[width = 1.\textwidth]{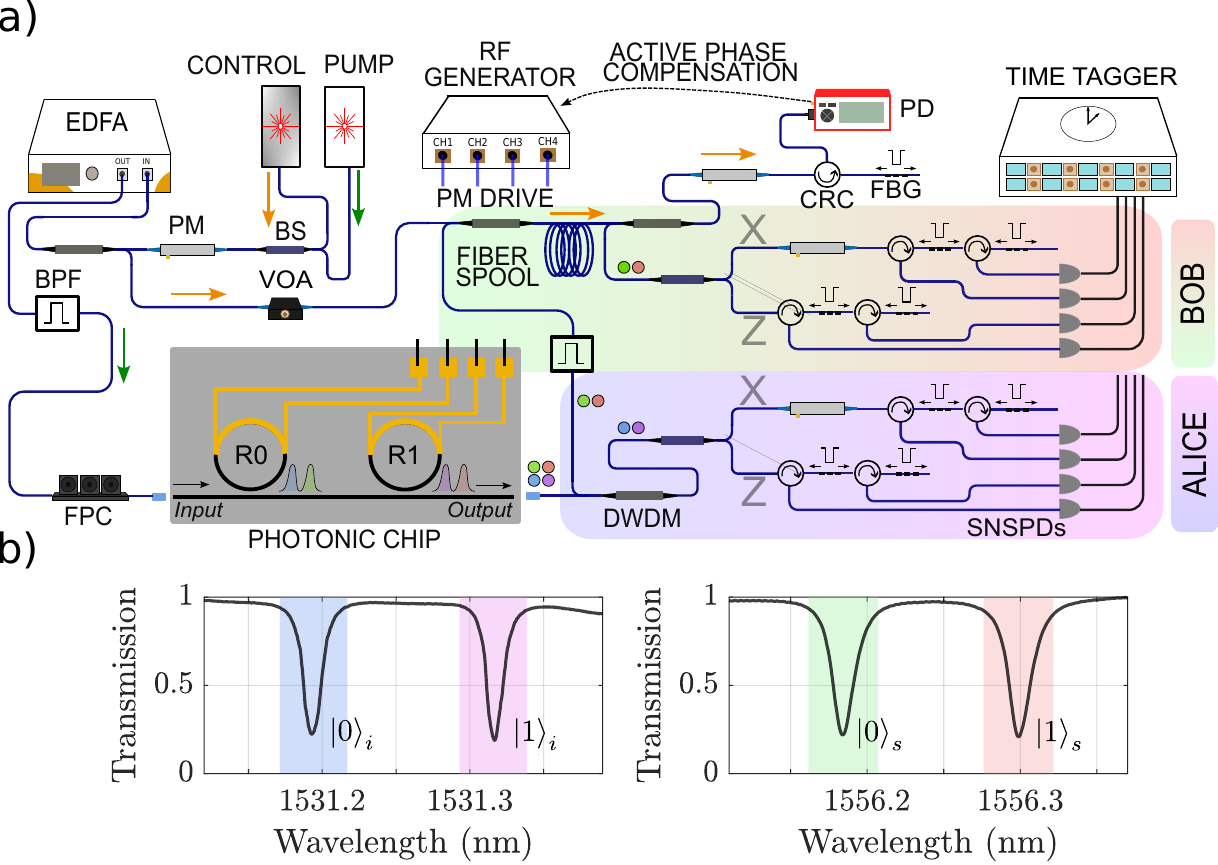}
    \caption{(a) Simplified sketch of the experimental setup. The direction of the pump(control) laser is indicated with a green(orange) arrow. Optical connections are shown in blue.  The frequency-bins of the Alice's photon (idler) are indicated by a blue and a violet filled circle, while those directed to Bob (signal) by a green and a red filled circle. BS: beamsplitter, EDFA: Erbium Doped Fiber Amplifier, VOA: Variable Optical Attenuator, FPC: Fiber Polarization Controller, BPF: bandpass filter, DWDM: Dense Wavelength Division Multiplexing, PM: electro-optic phase modulator, CRC: circulator, FBG: fiber Bragg grating, SNSPD: superconducting nanowire single photon detector, PD: photodiode. The layout of the chip shows the two resonators R0 and R1 (black), and the placement of the metallic heaters (yellow). (b) Spectral response of the device at the idler (left) and signal (right) wavelengths. The encoding of the signal and idler qubits is shown with the same colors used in panel (a).}
    \label{Fig_1}
\end{figure*}
A simplified version of the experimental setup is shown in Fig.\ref{Fig_1}(a). A tunable and continuous wave pump laser at $1543.69$ nm is sent through an electro-optic phase modulator (PM) driven by an RF signal at a frequency of $15$ GHz. Approximately three mutually coherent comb lines of equal intensity are created as a result of the phase modulation. After amplification and removal of the background noise, the pump is coupled to a silicon chip by using a lensed fiber. An on-chip inverse taper with a $160$ nm tip allows us to achieve a coupling loss of $3.5$ dB/facet. The pump excites two nominally identical microring resonators (labelled as R0 and R1 in Fig.\ref{Fig_1}(a)), which are coupled to the same bus waveguide. 
Metallic heaters are placed on the top of the rings and allows to thermally tune the resonance wavelength. More details on the sample geometry and fabrication are provided in \cite{clementi2023programmable,borghi2023reconfigurable}. The spectral response of the device is shown in Fig.\ref{Fig_1}(b), which highlights the set of resonances used 
to generate the signal and the idler entangled photon pairs by SFWM. The resonators have a free spectral range of $\sim 524$ GHz and a loaded quality factor of $10^5$. Each ring is actively locked to a different comb line of the pump laser using a Field Programmable Gate Array (FPGA) that controls the heater currents. A distinct dither tone is applied to each ring and revealed by homodyne detection to minimize the pump transmission. The locking allows to maximize the pair generation rate and to keep it stable over time.
The coherent driving of the two resonators allows the signal-idler pair to be generated either in R0 or in R1 in the corresponding set of resonances ($\ket{0}_s\ket{0}_i$ for R0, $\ket{1}_s\ket{1}_i$ for R1, see Fig.\ref{Fig_1}). We adjusted the relative intensity of the comb lines of the pump laser to generate the maximally entangled Bell state $\ket{\Psi^{+}} = \left (1/\sqrt{2} \right ) \left (\ket{0}_s\ket{0}_i+\ket{1}_s\ket{1}_i \right)$. A lensed fiber is used to couple light out of the chip, and a DWDM filter splits the signal and the idler photon into different paths. From now on, we will refer to the idler at $\sim 1531.3$ nm as the Alice's photon, while to the signal photon at $\sim 1556.3$ nm as the Bob's photon. The residual pump is removed in both paths using bandpass filters with high-extinction ratio. 
A fiber spool is then introduced into Bob's path. The length of the spool will be varied in the transmission tests discussed in Section \ref{sec:secure_key_rate} until a maximum of $26$ km.
We exploited $50/50$ beamsplitters to execute
passive basis selection among the $Z=\{\ket{0},\ket{1}\}$ and $X = \{ \ket{+} = (1/\sqrt{2})(\ket{0}+\ket{1}),\ket{-} = (1/\sqrt{2})(\ket{0}-\ket{1})\}$-basis. The measurement in the $Z$-basis is performed by using two fiber Bragg gratings (FBG, $12.5$ GHz of bandwidth) to reflect the $\ket{0}_{s(i)}$ and the $\ket{1}_{s(i)}$ frequency-bin into two superconducting nanowire single photon detectors (SNSPD) ($85\%$ of detection efficiency). The measurement in the $X$ basis is performed by mixing the $\ket{0}_{s(i)}$ and $\ket{1}_{s(i)}$ frequency-bins by using two PM \cite{clementi2023programmable}. A time tagging electronics is used to detect and record the arrival times of the signal and idler photons. The post-processing of the timestamps allows us to unambiguously identify the events of all the sixteen mutually exclusive outcomes, corresponding to the different Alice's and Bob's states combinations. It is worth to note that while the setup in Fig.\ref{Fig_1}(a) shows the use of eight SNSPDs, in the actual experiment we implemented only six detectors. We reduced the number of resources by using two detectors to perform both $X$ and $Z$-basis measurements, which are distinguished by introducing optical delays after passive basis selection. However, this scheme is equivalent to that shown in Fig.\ref{Fig_1}(a) which uses eight detectors, and that we adopted for ease of understanding. The detailed scheme of the actual experimental setup and the analysis of timestamps  is reported in detail in Appendix A.
\section{Characterization of the photon pair sources}
\label{sec:source_performance_characterization}
\begin{figure*}[t!]
    \centering
    \includegraphics[width = 0.83\textwidth]{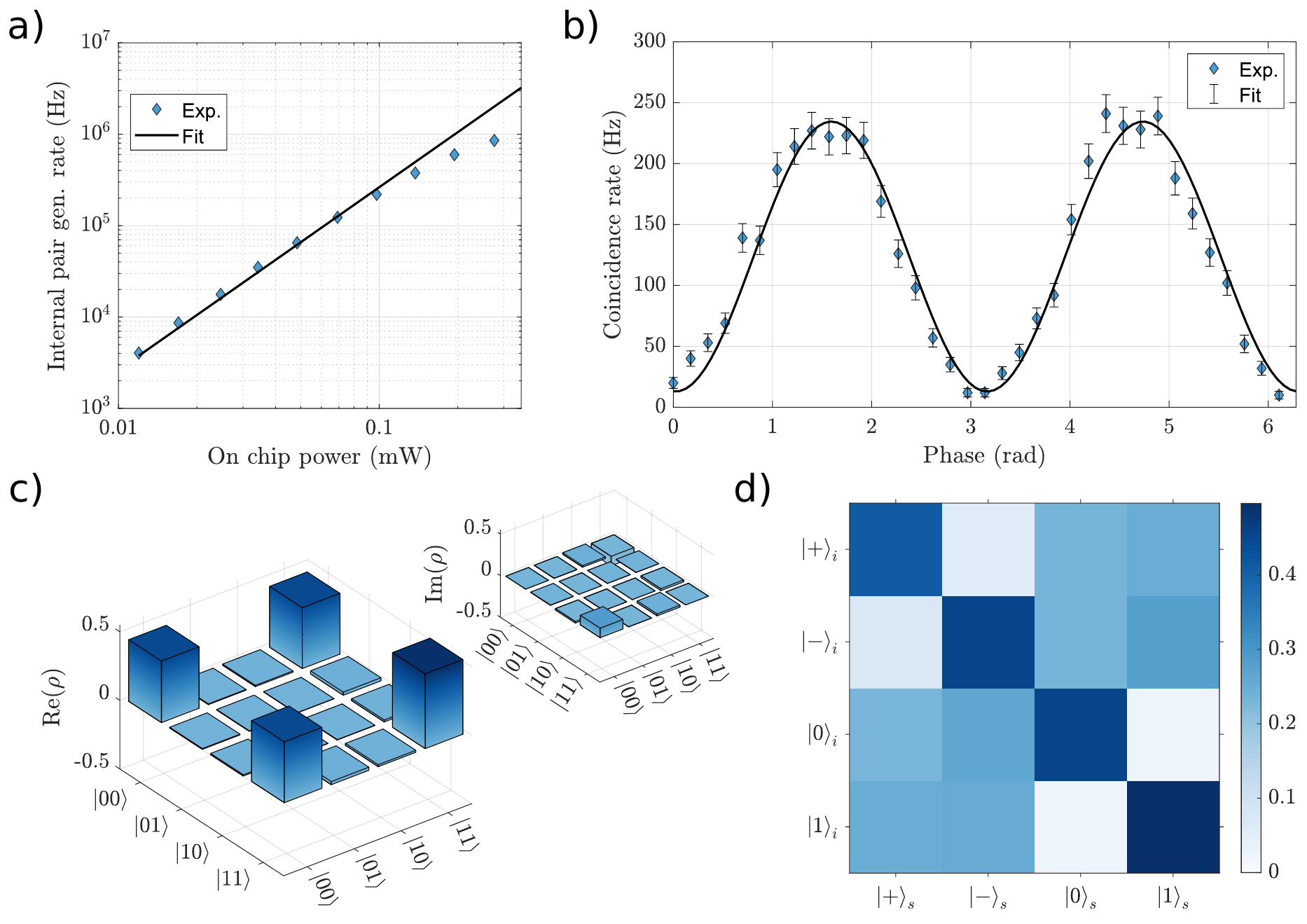}
    \caption{(a) On-chip pair generation rate of ring R0 for different pump powers $P$. The data is fit with the model $aP^2$ (solid black line) for $P<100\,\mu$W. (b) Bell fringe without any fiber spool applied to Bob's arm. Accidental counts are not subtracted. The solid black line is a sinusoidal fit of the data. (c) Reconstructed real and imaginary (top-right inset) part of the density matrix of the state at the output of the chip. (d) Correlation matrix of the probability outcomes on the two mutually unbiased basis sets $X_{s(i)}=\{\ket{+}_{s(i)},\ket{-}_{s(i)}\}$ and $Z=\{\ket{0}_{s(i)},\ket{1}_{s(i)}\}$ for the signal(idler) photon.}
    \label{Fig_2}
\end{figure*}
We characterized the brightness of both photon pair sources and the level of background noise, which will impact the secure key rate and the QBER discussed in Section \ref{sec:secure_key_rate}. The brightness is extracted by measuring the coincidence rate at the detector for different input pump powers, and by exciting only one source at a time. Coincidences are recorded within a window of $700$ ps.  This is shown in Fig.\ref{Fig_2}(a) for ring R0 (results for R1 are very similar and not shown). The on-chip generation rate is obtained from the raw coincidence rate at the detectors by correcting for the total losses from generation to detection. The coincidences are fit with a quadratic function of the pump power up to $100\,\mu W$, from which a brightness of $27(1)\,\textrm{MHz}\cdot\textrm{mW}^{-2}$ is extracted. The deviation from a quadratic behavior for input powers $>100\,\mu W$ is due to two-photon absorption and induced free-carrier absorption in silicon \cite{grassani2015micrometer}. From now on, in all the experiments we set the pump power of each comb line to $400\,\mu W$, corresponding to a on-chip pair generation rate of $\sim 0.7\,\textrm{MHz}$ and a measured coincidence to accidental ratio (CAR) of $\sim 20$ \cite{grassani2015micrometer}.\\  Another relevant metric that we extracted is the visibility of two-photon interference from the pairs generated by R0 and R1. The sources are simultaneously excited and their pair generation rate is equalized by regulating the relative intensity of the comb lines of the pump laser. We used  the Alice and Bob PM to project their respective photons along the equator of their Bloch sphere, i.e. we performed the local projector $\left ( 1/\sqrt{2}\right) \left ( \bra{0}_{s(i)}+e^{i\theta_{s(i)}}\bra{1}_{s(i)} \right )$, with $\theta_s = \theta_i=\theta$. We swept $\theta$ by changing the RF phase driving both PM, and recorded the two-photon interference fringe shown in Fig.\ref{Fig_2}(b). The raw visibility (without background subtraction) is $89.5(2)\%$, which increases to $93.3(2)\%$ by correcting for the accidental counts. 
The visibility is limited by the distinguishability between photons generated by R0 and R1 \cite{borghi2023reconfigurable}. To fully characterize the state, we also performed quantum state tomography using a complete set of projectors (the tensor product of the eigenvectors of three Pauli operators  $Z$, $X$ and $Y$ of the signal/idler qubits), and reconstructed the density matrix of the state using a standard maximum likelihood approach \cite{borghi2023reconfigurable}. The density matrix, shown in Fig.\ref{Fig_2}(c), has an Ulhman fidelity of $0.941(2)$ with the target $\ket{\Psi^{+}}\bra{\Psi^{+}}$.\\
The performance of the Alice and Bob receiver is assessed by performing correlation measurements in the $Z$ and $X$-basis. Ideally, the measurement outcomes should be perfectly correlated when the two parties choose the same basis ($XX$ and $ZZ$ correlation subspaces), and completely random when they choose different basis ($XZ$ and $ZX$ correlation subspaces). Randomness in the outcome is ensured by the fact that the two basis are mutually unbiased. The measured correlation matrix $T_{\textup{exp}}$ is shown in Fig.\ref{Fig_2}(d). 
The probabilities are normalized in each of the four ($XX,XZ,ZX,ZZ$)  $2\times2$ subspaces. We evaluated the fidelity $\mathcal{F}$ with the ideal correlation matrix $T_{\textup{th}}$ as
\begin{equation}
    \mathcal{F} = \frac{\textrm{Tr}(T_{\textup{exp}}^{\dagger}T_{\textup{th}})\textrm{Tr}(T_{\textup{th}}^{\dagger}T_{\textup{exp}})}{\textrm{Tr}(T_{\textup{exp}}^{\dagger}T_{\textup{exp}})\textrm{Tr}(T_{\textup{th}}^{\dagger}T_{\textup{th}})},
\end{equation}
where $T^{\dagger}$ is the matrix transpose. The fidelity between $T_{\textup{exp}}$ and $T_{\textup{th}}$ is $0.987(1)$, while the fidelities $\mathcal{F}_{ij}$ ($i=\{X,Z\},j=\{X,Z\})$ of the four subspaces are \mbox{$\mathcal{F}_{XX} = 0.975(2)$}, \mbox{$\mathcal{F}_{ZZ} = 0.995(2)$}, \mbox{$\mathcal{F}_{XZ} = 0.994(2)$} and \mbox{$\mathcal{F}_{ZX} = 0.997(2)$}. 

\section{Secure key rate analysis for different lengths of fiber spool}\label{sec:secure_key_rate}
\begin{figure}[h!]
    \centering
    \includegraphics[width=\linewidth]{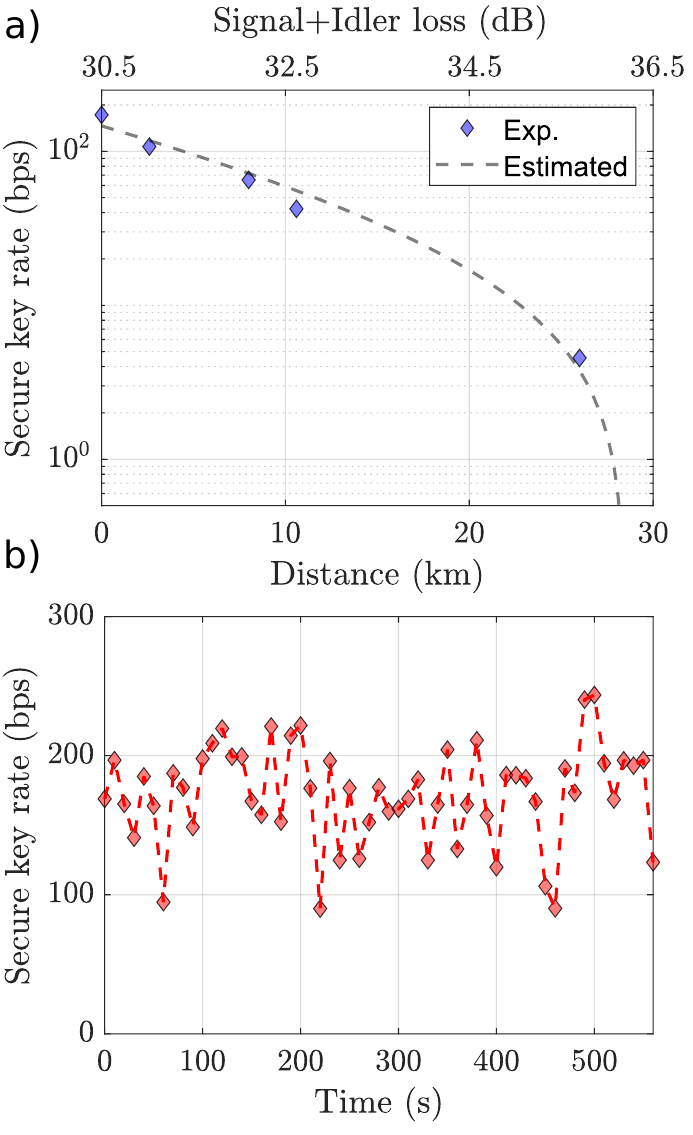}
    \caption{(a) Secure Key rate as a function of distance (length of the fiber spool). The blue points represent the experimental data. The dotted line represents the estimated SKR using Eq.(\ref{eq:skr}). The total loss include that related to the fiber spool and the loss from the chip output to the detectors (averaged over all the possible signal-idler paths leading to a coincidence event).  (b) Secure key rate as a function of time without the fiber spool inserted in the Bob path. 
    }
    \label{Fig_3_skr}
\end{figure}

We adopt the BBM92 protocol for secure key exchange, a variation of the BB84 protocol that employs entangled photons \cite{bennett1992quantum}, and calculate 
 the lower bound for the secure key rate under the infinite key approximation, which is given by \cite{ma2007quantum} 
\begin{equation}\begin{split}
    \textrm{SKR} \geq [1 - H_2(\varepsilon_Z)  f - H_2(\varepsilon_X)] \cdot S \, R_{r} \, \alpha \, \eta.
    \label{eq:skr}
\end{split}\end{equation}
Here, $H_2(x)$ is the entropy of the binary variable \textit{x}, $\varepsilon_Z(\varepsilon_X)$ is the QBER in the Z (X) basis, and $f>1$ is the error reconciliation efficiency. The term $H_2(\varepsilon_Z) f$\ represents the number of bits disclosed during the error correction protocol, so discarded from the total key length estimation. The term $H_2(\varepsilon_X)$ keeps into account the bits lost during the privacy amplification, necessary to minimize the correlation between Alice’s and Bob’s keys with the key owned by a potential eavesdropped (Eve). $S$\ represents the sifting ratio, i.e. the amount of bits discarded during the basis reconciliation process, $R_r$\ is the qubit generation rate, $\alpha$\ is the channel attenuation and $\eta$\ is the detectors' efficiency.\\
We evaluate the QBER and the SKR for fiber spools of different lengths, which are 0 km (no spool), 2.6 km, 8 km, 10.6 km and 26 km. The QBER in both the X and Z basis is listed in Table \ref{Table:qbers}.
We see that $\varepsilon_X>\varepsilon_Z$ for all the investigated fiber lengths. This is expected because the measurements in the X-basis are phase-sensitive, and errors are also introduced by the imperfect indistinguishability of the sources. In contrast, the QBER in the Z basis is primarily affected by the CAR, which remain stable as the photons travel through different fiber lengths without undergoing dephasing. The impact of dark counts from the detectors ($<120$ Hz) on the QBER is minimal because the associated coincidence rate is \mbox{$\sim10^{-3}$ Hz}. 
Figure \ref{Fig_3_skr}(a) shows the SKR achieved for the different lengths of the inserted fiber spool. 
Successful SKR extraction is demonstrated up to a fiber spool length of $26$ km, for which $\textrm{SKR}\ge4.5\,\textrm{bps}$. The propagation distance is limited by the (fixed) high losses of the Alice and Bob receiving stages. We did not noticed any impairment induced by the transmission in the fiber (for example fiber dispersion and/or spontaneous Raman scattering). Strategies for reducing the loss are discussed in Section \ref{sec:discussion}. 
The experimentally measured secure key rate aligns well with the expected results (dashed curve in Fig.\ref{Fig_3_skr}a), which are simulated by taking into account the source generation rate, the losses and the detector efficiencies. Up to $2.6$ km, the SKR is also stable over time, as shown in Fig.\ref{Fig_3_skr}(b) in the case where no fiber spool inserted into Bob's path.

\begin{table}
\begin{tabular}{c|c|c}
Fiber length (km) & $\varepsilon_Z$ & $\varepsilon_X$  \\ 
\hline
0 & 0.049(2) & 0.13(1)   \\
2.6 & 0.054(2) & 0.14(1)   \\
8 & 0.057(2) & 0.138(8)   \\
10.6 & 0.057(2) & 0.14(1) \\
26 & 0.065(3) & 0.143(8)
\end{tabular}
\caption{QBER measured in the X and Z basis for different length of fiber spools inserted in the Bob arm.}
\label{Table:qbers}
\end{table}

\section{Active phase-drift compensation}
\label{sec:active_phase_compensation}
\begin{figure*}[t!]
    \centering
    \includegraphics[width = \textwidth]{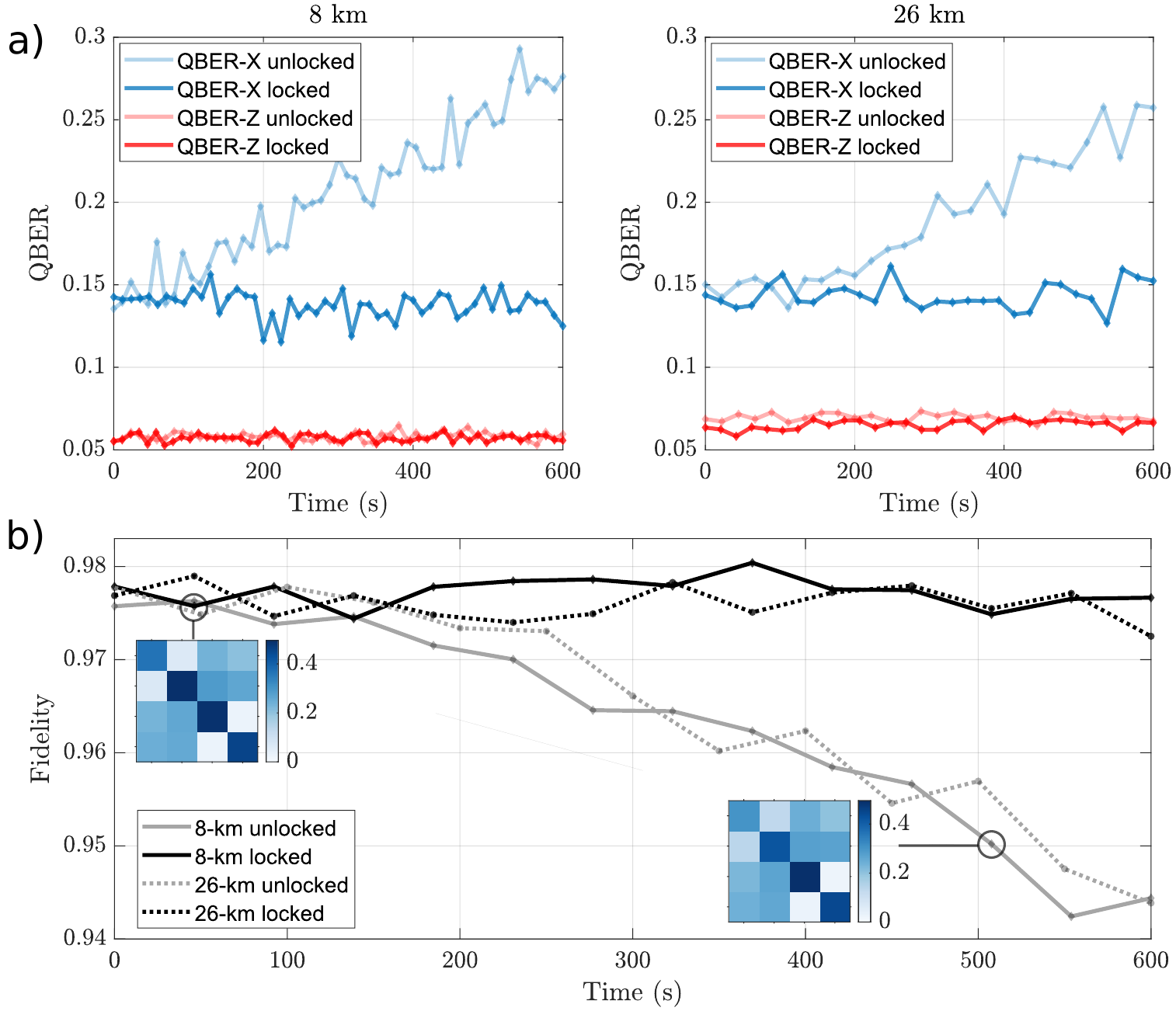}
    \caption{(a) QBER variations on the $X$ and $Z$ basis as a function of time with (locked) and without (unlocked) active phase compensation. In left-panel, the signal photon propagates along $8$ km of fiber spool, while in the right panel along $26$ km. (b) Fidelity between the experimental and the theoretical correlation matrix of the outcomes between the $X$ and $Z$ basis as a function of time. The black(gray) lines indicate the fidelity with the active phase compensation turned on(off). The two insets show the correlation matrix at two different times.}
    \label{Fig_4}
\end{figure*}
We assessed the stability of the QBER on the $X$ and $Z$ basis over time for all the different lengths of fiber spool shown in Fig.\ref{Fig_3_skr}. Up to $2.5$ km, the QBER on both basis does not vary significantly during a monitoring time of $600$ s, with an root mean square variation $<6\%$. At longer propagation lengths, we systematically observed an increase of the $\varepsilon_X$ with time, with no variations of the $\varepsilon_Z$. This trend is shown in Fig.\ref{Fig_4}(a) for a fiber length of $8$ km and $26$ km. Consequently, the fidelity $\mathcal{F}$ between the experimental and the ideal correlation matrix decreases over time, as shown in Fig.\ref{Fig_4}(b). From an initial value of $\mathcal{F}\sim 0.98$, the fidelity drops to a final value of $\mathcal{F}\sim 0.94$ after $\sim 600$ s. By comparing the initial and the final correlation matrices (insets in Fig.\ref{Fig_4}(b)), we notice that only the elements in the $XX$ subspace are changed. In particular, the off-diagonal components increased, indicating that the state at the Alice and Bob's modulators is no longer $\ket{\Psi^{+}}$ but gets rotated in time by some angle $\theta$, i.e $\ket{\Psi^{+}} \rightarrow \left (1/\sqrt{2} \right ) \left (\ket{00}+e^{i\theta}\ket{11} \right)$. Importantly, this rotation does not affect the correlations on the $ZX$ and $ZZ$ subspaces. We attributed the rotation to a time-varying variation of the optical path between the $\ket{0}_s$ and $\ket{1}_s$ frequency-bin states during the propagation along the fiber spool. Indeed, if the signal and idler photon are sent to two different fibers of length $L_s$ and $L_i$, with $L_s\gg L_i$, then $\theta = 2\pi \Delta \nu \delta[n(L_s-L_i)]/c\sim 2\pi \Delta \nu \delta[nL_s]/c$, where $\delta[nL_s]$ is the variation of the optical path $nL_s$, $n$ is the effective index of the optical fiber and $c$ is the speed of light. The phase change is then equivalent to that experienced by a photon of frequency $\Delta\nu$, which in our experiment corresponds to a wavelength in the material of $\sim 1.5$ cm. Thus, an optical path variation of the order of few cm is sufficient to transform the state from $\ket{\Psi^{+}}$ to $\ket{\Psi^{-}}$, modifying the measurement outcomes on the $XX$ subspace from being perfectly correlated to be completely anti-correlated. To the first order, $\delta[nL_s] = \delta[n]L_s+n\delta[L_s]$, i.e., both refractive index variations and changes in the physical length of the fiber affect the optical path length. The slow drift of $\theta$ (hence of $\varepsilon_X$ and of $\mathcal{F}$) seen in Fig.\ref{Fig_4} suggests that such variations could be induced by temperature fluctuations of the environment. To verify this hypothesis, we conducted an experiment in which the temperature of a $26$ km fiber spool is changed, and the optical path length $\Delta L_{\textup{eff}}=\delta[nL_s]$ is continuously monitored from the time of flight of a ps laser pulse propagating in the spool, which is measured by a fast APD with $40$ ps of timing jitter. By correlating $\Delta L_{\textup{eff}}$ with the temperature variation, which is measured by a thermistor placed in direct contact with the spool, a linear relationship is found, with a slope of $1.01(1)\,\frac{\textrm{cm}}{\textrm{km}{}^{\circ}C}$. In addition, we continuously monitored the temperature of the laboratory for $10$ hours, from which we evaluated an average variation of $\sim 0.03(1) {}^{\circ} C$ every $600$ s. From this we extrapolated an average variation of the optical path of $\sim 0.03$ cm ($\theta\sim 0.1$ rad)  per km of fiber every $600$ s. These results indicate that the random temperature fluctuations of the environment are sufficient to cause a change of $\theta$ of the order of few radians for fiber spools of tens km of length, which justify the degradation of the $\varepsilon_X$ over time observed in Fig.\ref{Fig_4}(a). \\
The phase drift must be compensated to guarantee stable data transmission with a low QBER. To this goal, we developed a technique that tracks $\theta$ and actively compensates any variation by rotating the $X$-basis of the Bob's photon by the same amount. This is accomplished with few extra components, shown in Fig.\ref{Fig_1}(a). A weak ($\sim -45$ dBm) control beam at $1544.5$ nm is multiplexed with the pump laser and passed into the PM. 
The control beam is then demultiplexed and injected into the fiber spool, where it co-propagates along with the signal photon. A DEMUX extracts the control beam after the fiber spool and a PM (denoted as control PM) is used to coherently mix the three frequency bins of the control comb. A FBG with a passband of $12.5$ GHz is used to filter the baseband signal, which is detected by a photodiode. This scheme effectively realizes a multi-mode Mach-Zehnder interferometer in the frequency domain which is sensitive to $\theta$  (see Appendix B for more details).
We repeatedly swept the phase of the RF signal of the control PM, and from the recorded inference fringe we tracked the phase drift $\theta$ between the $\ket{0}_s$ and $\ket{1}_s$ frequency-bins. This phase is then added to the RF modulator of the signal, thus compensating the rotation of the state in the Bloch sphere by applying an analogous rotation to the Bob's reference frame. The phase correction is provided every $\sim 2$ s, which is sufficiently fast to compensate the slow phase drift caused by the temperature variation of the environment. No effort has been made to improve the running time of the whole procedure. As shown in Fig.\ref{Fig_4}(a), when the active phase compensation mechanism is turned on, the slow drift of the $\varepsilon_X$ over time is eliminated and its value remains constant around $15\%$ for every investigated fiber length. At the same time, the fidelity $\mathcal{F}$ lies always above $0.975$ (see Fig.\ref{Fig_4}(b)). The SKR reported in Fig.\ref{Fig_3_skr} and the QBER values listed in Table \ref{Table:qbers} are all calculated with the active phase compensation system turned on.  We found an almost perfect (Pearson correlation coefficient of $0.995$) correlation between the phase variation $\theta$ and the temperature change of the fiber spool, with a slope of $19(1)\,\textrm{deg}\cdot\textrm{km}^{-1}\textrm{GHz}^{-1}{}^{\circ}\textrm{C}^{-1}$ (see Appendix B). By assuming a thermo-optic coefficient of $1.1\cdot10^{-5}\,\textrm{C}^{-1}$ and a thermal expansion coefficient of $5\cdot10^{-7}\,\textrm{C}^{-1}\textrm{m}^{-1}$ for the silica fiber \cite{gao2018investigation}, the predicted slope is $\sim 14\,\textrm{deg}\cdot\textrm{km}^{-1}\textrm{GHz}^{-1}{}^{\circ}\textrm{C}^{-1}$, very close to the experimental value. The higher slope found in the experiment could be due to the under-estimation of the temperature variation of the fiber, which is sensed by the thermistor in the outermost part of the spool.    
\section{Discussion\label{sec:discussion}}
The results of Sections \ref{sec:secure_key_rate} and \ref{sec:active_phase_compensation} are a proof-of-concept demonstration of entangelement-based QKD with frequeny-bin encoding. However, despite the high-brightness of the source, the SKR at the detector is not competitive with state-of-the art works using time-bin or polarization encoding \cite{fitzke2022scalable,appas2021flexible}. The  main limiting factor are the high-losses from pair generation to detection. They result from the poor coupling efficiency with the chip and from the high  insertion losses of the optical components, such as the pump filters, the PMs and the connector loss. Indeed, on average losses from generation (excluding the losses due to the fiber spool) to detection are $\sim 10$ dB for Alice and $\sim 24$ dB for Bob (see Table \ref{Table_2} in Appendix A). The higher losses of the Bob's path are due to the use of lossy tunable bandpass filters and from the MUX/DEMUX of the control laser beam. In addition, the projectors on the $X$-basis use post-selection \cite{lu2020fully}, which introduces an excess loss of $\sim 2.2$ dB (included in the total insertion loss reported before).\\
However, losses can be greatly reduced by optimizing the geometry of the device. For example, choosing a resonator FSR that is a multiple of the standard ITU grid ($100-200$ GHz), would allow us to use MUX/DEMUX 
with low loss ($<0.5$ dB) and high-exctintion ratio, which could replace our lossy tunable filters. Coupling loss could be brought below $1$ dB by using optimized inverse tapers \cite{du2024demonstration}. 
Losses induced by post-selection represent a major obstacle. Indeed, Hadamard gates without post-selection have been demonstrated using a combination of waveshaper and electro-optic modulators, but this solution has an insertion loss of $\sim10$ dB when implemented with standard fiber-optic components \cite{henry2023parallelizable}. Nevertheless, on chip-waveshapers with GHz resolution have been recently reported \cite{cohen2024silicon}, and there is hope that hybrid-platforms embedding PM could decrease the loss of aggregate components  \cite{wang2023integrated}. In light of these considerations, it is realistic to expect that losses between $5-7$ dB could be reached with a small device optimization, which would increase the coincidence rate at the detectors by more than two orders of magnitude. \\
In parallel, the number of pairs delivered per second could be increased by frequency multiplexing. One could use the $\ket{0}_{s(i)}$ and $\ket{1}_{s(i)}$ frequency-bins of multiple FSR to encode multiple qubits and process them in the same spatial mode \cite{henry2023parallelizable,lu2018electro}. For example, in \cite{henry2024parallelization} the authors report the simultaneous manipulation of $17$ frequency-bin qubits with an offset frequency of $21$ GHz and separated by a guard-band of $\sim 40$ GHz. \\
The high-dimensional encoding of frequency-bin qudits could also be harnessed to increase the information content per photon and the resilience against errors which are introduced by the transmission loss and imperfections of the receiver \cite{cozzolino2019high}. In \cite{borghi2023reconfigurable} some of the authors demonstrated the generation of reconfigurable frequency-bin entangled qudits up to a dimension of four by scaling the number of rings in this work from two to four, with the possibility to directly manage the pump power on chip and to choose the frequency-bin separation. The use of multiple rings of high-finesse in place of a single resonator of large-volume allows the source brightness to be increased, thereby decreasing the input pump power required to achieve a target pair generation rate. This is beneficial for reducing the nonlinear noise constituted by photon pairs generated in other parts of the chip \cite{burridge2022quantifying}. In addition, the use of multiple rings allows the group of resonances encoding the qudit state, which are typically separated by few tens of GHz, to be naturally isolated from the subsequent FSR, providing a guard-band that prevents mixing with the other comb lines. This is an important advantage over the use of a single resonator of low-finesse, where guard-band modes must be filtered before applying electro-optic modulation.\\    
A separate issue that we did not address in this work is the RF synchronization of the PM at the two remote nodes where the Alice and Bob receivers are located. In our proof-of-concept demonstration, we used a fiber spool to emulate the link distance between the two parties, which are physically located in the same place. This allowed us to drive the two PM using standard RF cables that carry the signal from a common oscillator. In a real implementation Alice and Bob are  physically separated, which implies that two remote oscillators must be synchronized to sub-ps precision to achieve stable coherent operation. An important step towards this goal was taken in \cite{chapman2024quantum}, where the authors demonstrated the distribution of a $19$ GHz clock  over $5.5$ km using a radio-over-fiber (RFoF) system, with a drift of less than $0.5$ ps on $30$ minutes of operation. In this demonstration, the RF tone driving the PM at the sender node is intensity-encoded and multiplexed together with the quantum signal towards the end node, where it is demultiplexed and used to drive the end-node PM after electro-optic conversion and amplification. This strategy has been proved to be robust against thermal fluctuations, and could be used to simultaneously distribute the clock and to eliminate the phase drift between the frequency bins.\\

\section{Conclusions}
\label{sec:conclusions}
We presented the first demonstration of entanglement-based quantum key distribution based on frequency-bin encoding of photonic qubits. Two silicon microring resonators of high-finesse are used to generate a maximally entangled Bell state at a rate of $0.7$ MHz. We performed passive basis selection between the mutually unbiased $Z$ and $X$ sets and simultaneously monitored all the possible sixteen mutually exclusive outcomes. Transmission tests were conducted up to $26$ km of fiber-length with a secret key rate $\ge4.5$ bit/s. We demonstrated that this type of encoding is robust against bit-flip errors, but it  requires an active phase compensation system to correct for phase drifts between different frequency-bins caused by temperature fluctuations in the environment. In our tests, the drift has time-scale variations of a few seconds and has been quantified to be \mbox{$19(1)\,\textrm{deg}\cdot\textrm{km}^{-1}\textrm{GHz}^{-1}{}^{\circ}\textrm{C}^{-1}$}. We developed a real-time feedback loop that recovers the phase drift and eliminates errors induced  at the receiver by rotating the measurement base. The relatively low  secret key rate achieved is limited by the high losses of our implementation, but could be increased by more than two orders of magnitude by optimizing the geometry of the device. The use of wavelength division multiplexing and of high-dimensional encoding is a natural extension of the presented architecture, especially since multiple single-qubit gates can be parallelized in the frequency domain without increasing the hardware resources, and would allow to achieve higher bit rates and increased noise resilience.

\subsection*{Funding and acknowledgements}
The authors acknowledge Federico Andrea Sabattoli, Houssein El Dirani, Laurene Youssef, Camille Petit-Etienne, Erwine Pargon, and Corrado Sciancalepore for the design and fabrication of the sample. The device has been fabricated at CEA-LETI (Grenoble) on a 200 mm silicon-on-insulator wafer manufactured by SOITEC (Bernin), and is part of a previous collaboration. 
D.B acknowledges the support of Italian MUR and the European Union - Next Generation EU through the PRIN project number F53D23000550006 - SIGNED. M.B., M.G and M.L. acknowledge the PNRR MUR project PE0000023-NQSTI. N.T acknowledges the HyperSpace project (project ID 101070168). 
D. Bacco acknowledges the support of the European Union (ERC, QOMUNE, 101077917, the Project SERICS (PE00000014) under the MUR National Recovery and Resilience Plan funded by the European Union - NextGenerationEU. D.R. acknowledges the Project QUID (Quantum Italy Deployment) funded by the European Commission in the Digital Europe Programme under the grant agreement No 101091408.

\section*{Appendix A: detailed description of the experimental setup and of the timestamp analysis}
\label{sec:experimental_setup}

\begin{figure*}[t!]
    \centering
    \includegraphics[width = 1 \textwidth]{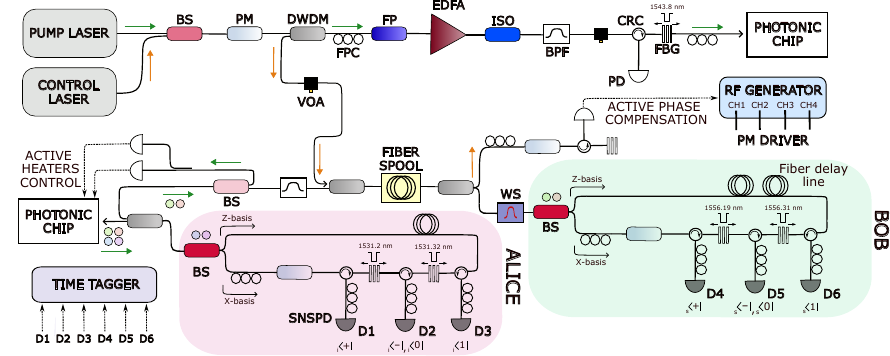}
    \caption{Scheme of the experimental setup. The direction of the pump (control) laser is indicated with a green(orange) arrow. The frequency-bins of the Alice’s photon (idler) are indicated by a blue and a
    violet filled circle, while those directed to Bob (signal) by a green and a red filled circle. Alice's (Bob's) passive basis selection measurement stage is highlighted in pink (green).  BS: beamsplitter, PM: electro-optic phase modulator, DWDM: Dense Wavelength Division Multiplexing, FPC: Fiber Polarization Controller, FP: Fiber Polarizer, EDFA: Erbium Doped
Fiber Amplifier, ISO: fiber Isolator, BPF: bandpass filter,  VOA: Variable Optical Attenuator, PD: photodiode,
CRC: circulator, FBG: fiber Bragg grating, SNSPD: Superconducting Nanowire Single Photon Detector, WS: Waveshaper. }
    \label{Fig_Setup}
\end{figure*}
\begin{table*}[t!]
\makebox[\linewidth]{
\begin{tabular}{{c|c|c|c|c}}
\textbf{Detectors pair }& \textbf{Alice's basis } & \textbf{Bob's basis } & \textbf{Projector} & \textbf{Arrival time  } \\
\hline 
\multicolumn{1}{l|}{} & \multicolumn{1}{l|}{} & \multicolumn{1}{l|}{} & \multicolumn{1}{l|}{} & \multicolumn{1}{l}{} \\
D1 - D4& X & X& $_{is}\bra{++}$ & 0 \\
D1 - D5& X & X& $_{is}\bra{+-}$ & 0 \\
D1 - D5& X & Z & $_{is}\bra{+0}$ & 2$\tau$ \\
D1 - D6 & X & Z& $_{is}\bra{+1}$ & 2$\tau$\\
D2 - D4          & X             & X           & $_{is}\bra{-+}$ & 0             \\
D2 - D4          & Z             & X           & $_{is}\bra{0+}$ & -$\tau$     \\
D2 - D5          & X             & X           & $_{is}\bra{--}$ & 0             \\
D2 - D5          & X             & Z           & $_{is}\bra{-0}$ & 2$\tau$     \\
D2 - D5          & Z             & X           & $_{is}\bra{0-}$ & -$\tau$     \\
D2 - D5          & Z             & Z           & $_{is}\bra{00}$ & $\tau$      \\
D2 - D6          & X             & Z           & $_{is}\bra{-1}$ & 2$\tau$     \\
D2 - D6          & Z             & Z           & $_{is}\bra{01}$ & $\tau$      \\
D3 - D4          & Z             & X           & $_{is}\bra{1+}$ & -$\tau$     \\
D3 - D5          & Z             & X           & $_{is}\bra{1-}$ & -$\tau$     \\
D3 - D5          & Z             & Z           & $_{is}\bra{10}$ & $\tau$      \\
D3 - D6          & Z             & Z           & $_{is}\bra{11}$ & $\tau$     
\end{tabular}
}
\caption{List of the sixteen projectors that are simultaneously record during the experiment. For each projector, we specify which pair of detectors clicks and the difference in arrival time between Alice's (start) photon and Bob's (stop) photon. The delay $\tau$ is $\sim10$ ns.}
\label{Table_1}
\end{table*}

\begin{table*}
\centering

\begin{tabular}{c|c|c|c}
\textbf{\textbf{Alice/Bob Detector}} & \textbf{\textbf{Measurement basis}} & \textbf{\textbf{Projector}} & \textbf{\textbf{Loss without fiber spool (dB)}}  \\ 
\cline{1-4}
&  &  &   \\
D1 & X & $_{i}\bra{+}$ & 13.0(5)   \\
D2 & X & $_{i}\bra{-}$ & 13.9(5)   \\
D2 & Z & $_{i}\bra{0}$ & 8.4(5)   \\
D3 & Z & $_{i}\bra{1}$ & 9.1(5)  \\
D4 & X & $_{s}\bra{+}$ & 22.5(5)   \\
D5 & X & $_{s}\bra{-}$ & 21.7(5)   \\
D5 & Z & $_{s}\bra{0}$ & 16.7(5) \\
D6 & Z & $_{s}\bra{1}$ & 16.6(5) 
\end{tabular}
\caption{Losses measured from out of the chip to the detectors for each of the possible Alice and Bob projectors. Losses do not include the propagation in the fiber spool in Bob's path, but consider detector efficiency and post-selection in the $X$-basis.}
\label{Table_2}
\end{table*}
A sketch of the complete experimental setup is shown in Fig.\ref{Fig_Setup}. A pump laser (Santec TSL-570) tuned at $1543.69$ nm and a control laser (Emcore) at $1544.53$ nm are combined through a $90/10$ polarization maintaining beamsplitter. The output is modulated by and electro-optic phase modulator (PM) driven by an RF signal at a frequency of $15$ GHz. The power of the RF signal ($\sim$ 23 dBm) is sufficient to create approximately three mutually coherent comb lines of equal intensity for the pump control laser. The RF generator (Anapico) is used to drive all the PM in the setup shown in Fig.\ref{Fig_Setup}.  A DWDM filter (Opneti CH$41$, $100$ GHz) separates the control laser from the pump. The first is attenuated to a power of \mbox{$\sim$ $-45$ dBm} using a VOA and sent to another DWDM filter, where it is combined with the Bob's photons. The pump is amplified by polarization mantaining EDFA (Keopsys CEFA-C-PB-LP). A fiber polarization controller and an in-line fiber polarizer are used to match the polarization of the EDFA. A high-power fiber isolator protect the EDFA amplifier from the back-reflected light. We used a tunable bandpass filter (Satec OTF tunable filter, bandwidth $\sim$ 0.3 nm), centered at the pump wavelength to remove the background noise.
Before coupling the pump laser to the photonic chip, one of the comb lines is removed by using a circulator and a Fiber Bragg Grating (FBG, from O/E-land) with a bandwidth of $12.5$ GHz. This procedure greatly reduces the generation of spurious photons into the bus waveguide.\\
The silicon photonic chip is composed by two identical microring resonators which are coherently and indipendently pumped by the two comb lines at the input of the chip.  
We used two lensed fibers with a spot size of $\sim 2.5\,\mu$m to couple light into and out of the photonic chip. On-chip inverse tapers allows us to achieve an insertion loss of $\sim\,7$ dB. The on-chip power is regulated using an electronic variable optical attenuator (Thorlabs).
After the chip, a DWDM filter (Opneti CH58, 100 GHz) is used to separate Alice's (idler) photons from residual pump and Bob's (signal) photons. Alice's photons are then sent to the passive basis selection stage. 
In the path of Bob's photons, we placed a $99/1$ beamspitter tap to monitor the transmitted light through the chip. This is used to generate the error signal at the input of a Field Programmable Gate Array (Red Pitaya STEMlab 125-14), which locks each ring to a different comb line of the pump laser by acting on the heater driver.  The 99$\%$ of light passes through a series of two bandpass filters (Semrock) that completely suppress the residual pump laser. Bob's photons are then combined to the control laser by using a DWDM filter (Opneti CH41, 100 GHz), and propagate in a fiber spool. We used fiber spools of different length in the transmission tests of Section IV of the main manuscript ($2.6$ Km, $8$ Km, $10.6$ Km and $26$ Km). At the output of the spool, the control laser is separated from the Bob's photons by a DWDM (Opneti CH41, 100 GHz) and sent to a PM that coherently mixes its comb-lines. 
A circulator and a FBG ($12.5$ GHz of bandwidth) are used to filter the base-band signal, which is detected by a photodiode. By sweeping the phase of the RF signal on the control PM we can extract the phase drift between $\ket{0}_{s}$ and $\ket{1}_{s}$ frequency-bins, as discussed in detail in Section V of the main manuscript and in Section \ref{sec:phase_drift_characterization}. 
Bob's photons pass through a waveshaper (II-VI 16000A) which is used as a bandpass filter at $1556.2$5 nm and with a bandwidth of $0.35$ nm. Finally, they are sent to the Bob's passive basis selection stage.\\
Alice's and Bob's projective measurement stages execute a random passive basis selection among the \mbox{$Z=\{\ket{0},\ket{1}\}$} and \mbox{$X = \{ \ket{+},\ket{-}\}$}-basis by using a $50/50$ fiber beasmplitter. Both stages allow to simultaneously record all the sixteen projective measurements by exploiting six superconducting nanowire single photon detectors (SNSPDs). 
The measurement in the Z-basis is performed by separating the $\ket{0}_{i(s)}$ from the $\ket{1}_{i(s)}$ frequency-bins by using a combination of FBGs ($12.5$ GHz of bandwidth) and circulators, and by directing the filtered frequency to separate SNSPDs, that we label D2 and D3 for Alice, D5 and D6 for Bob. Fiber polarization controllers are used to maximize the polarization-dependent detection efficiency.
The measurement in the X basis is executed by mixing the
$\ket{0}_{i(s)}$ and $\ket{1}_{i(s)}$ frequency-bins with a PM \cite{clementi2023programmable}. Next, the frequency-bins are filtered through FBGs and circulators, and directed to detectors D1 and D2 for Alice, D4 and D5 for Bob. The detection efficiency of the SNSPDs D1, D2, D3, D5 is $85\%$, while that of D4 and D6 is $73\%$. A time tagging electronics is used to extract coincidences between Alice and Bob's detectors though timestamps analisys.  Since D2 and D5 are used to perform both $X$ and $Z$-basis measurements, we introduced fiber delays ($\tau\sim 10$ ns and $2\tau\sim 20$ ns, see Fig.\ref{Fig_Setup}) to tell the basis choice from the relative arrival time between the Alice and Bob photon. The delay and the pair of detectors that clicks unambiguously identify the sixteen projective measurement, as detailed in Table \ref{Table_1}. Photons arriving at D1 and D4 have a relative delay equal to zero, and the delay is always referred to the arrival time of the idler photon.\\ 
The losses of each projective measurement are listed in Table \ref{Table_2}. They  account for the total transmission from the chip to detection, including the detector efficiency and the losses induced by post-selection. The insertion of fiber spools in the Bob's path increases the losses by $1.4(2)$ dB for a $2.6$ Km long spool, $2.0(2)$ dB for a $8$ Km long spool, $3.2(2)$ dB for a $10.6$ Km long spool and $5.0(2)$ dB for a $26$ Km long spool. 
\section*{Appendix B: Characterization of the thermal phase noise}\label{sec:phase_drift_characterization}
\begin{figure*}[t!]
    \centering
    \includegraphics[width = 0.78 \textwidth]{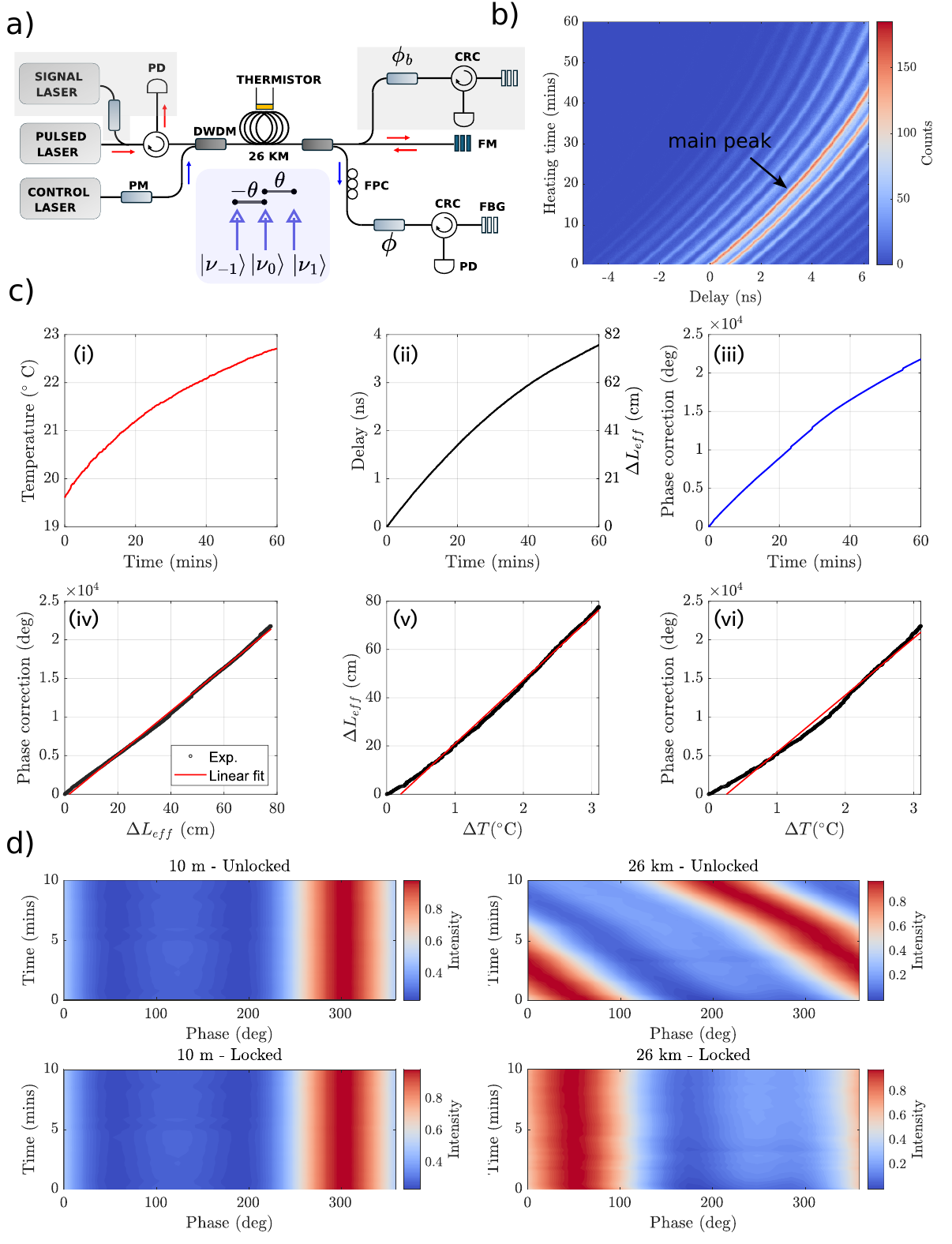}
    \caption{(a) Sketch of the experimental setup used to characterize the phase drift between the frequency-bins in a $26$-km fiber spool. The optical paths of the pulsed and control laser are indicated with a red and a blue arrow respectively. DWDM: Dense Wavelength Division Multiplexing, PM: electro-optic phase modulator, FM: Faraday mirror, FPC: fiber polarization controller, CRC: fiber optic circulator, FBG: fiber Bragg grating, PD: photodiode. The components highlighted in gray are used only when the active phase compensation system is turned on. (b) Stacked histogram showing the time evolution of the reflected pulse laser intensity as a function of the optical delay with the electronic trigger. The black arrow indicates the reference peak, while the sideband ripples are due to spurious reflections. (c) From (i) to (iii): temperature, optical delay and aggregated phase correction as a function of the cooling time of the fiber spool. (iv) Aggregated phase correction as a function of the optical path length variation $\Delta L_{\textup{eff}}$. (v) Optical path length variation (v) and aggregated phase correction (vi) as a function of the temperature change $\Delta T$ of the fiber-spool. (d) Evolution of the interference fringe of the control laser over time in a fiber length of $10$ m with the active phase compensation mechanism turned off (top panel) and on (bottom panel). The right panels refer instead to a $26$ km fiber spool.}
    \label{Fig_S1}
\end{figure*}
The relative phase $\theta$ between frequency-bins $\ket{0}$ and $\ket{1}$ propagating in a fiber of length $L$ is \mbox{$\theta = 2\pi\Delta\nu nL/c = 2\pi\Delta\nu L_{\textup{eff}}/c$}, where $n$ is the effective index of the fiber and $c$ the speed of light. Temperature variations of the environment change the optical path length $L_{\textup{eff}}=nL$ due to the thermo-optic effect and thermal expansion, thus imparting a phase drift between frequency-bins. We characterized and quantified this effect using the setup sketched in Fig.\ref{Fig_S1}(a). A pulsed laser with a duration $\sim\,1$ ps at a wavelength of $1550$ nm is sent into a fiber spool of length $26$ km and is reflected back by a Faraday mirror to double the propagation length. The reflected pulse is filtered using a circulator placed at the input of the spool, and is attenuated down to the single photon level before being detected by a single photon avalanche photodiode (SPAD) with $40$ ps of timing jitter. The signal from the SPAD is sent to a time tagging electronics which is synchronized with the $10$ MHz clock of the laser. This allows us to retrieve the time of flight of the pulse in the spool. The temperature of the latter is brought to $\sim 19.5\,{}^{\circ}$ C by using a refrigerator, and then allowed to warm up slowly to room temperature ($\sim 25\,{}^{\circ}$ C). The time of flight is continuously monitored during the heating process together with the temperature of the spool, which is sensed by a thermistor placed on the surface. Figure \ref{Fig_S1}(b) shows the time evolution of the reflected pulse as a function of the its optical delay $\tau$ with the electronic trigger. 
The delay $\tau_d$ and the temperature monotonically increase over time, as shown in Fig.\ref{Fig_S1}(b-c(i,ii)). Due to our choice of using the arrival time of the optical signal $t_s$ from the SPAD as the stop and the arrival time of the electronic trigger $t_t$ as the start, we have that $\tau_d = t_s-t_t$. Since $t_t$ is fixed, the trend shown in Fig\ref{Fig_S1}(b,c(ii)) indicates that the time of flight $\tau$, thus the optical path length $\Delta L_{\textup{eff}}=2v_g\tau_d$ ($v_g$ is the group velocity in the fiber) is increasing with temperature. This is expected because both the thermo-optic coefficient $\alpha_{to}=1.1\cdot 10^{-5}\,\textrm{C}^{-1}$ and the thermal expansion coefficient $\alpha_{exp}=5\cdot\,10^{-7} \textrm{C}^{-1}\textrm{m}^{-1}$  of the fiber are positive \cite{gao2018investigation}.\\
In order to correlate $\Delta L_{\textup{eff}}$ with the phase drift between the frequency bins, a control laser at frequency $\nu_0$ ($\lambda_0 = 1544.54$ nm) and power $\sim -45$ dBm is multiplexed with the pulsed laser before entering into the spool, and demultiplexed at the output before the Faraday mirror. The control laser at the input is phase modulated at $\Delta\nu = 15$ GHz using an electro-optic modulator (PM). The RF power is regulated to achieve a modulation index $\delta$ equal to $\sim 1.4$, which in the spectral domain corresponds to the transformation \mbox{$\ket{\Psi}\sim J_{-1}\ket{\nu_{-1}}+J_{0}\ket{\nu_{0}}+J_{1}\ket{\nu_{1}}$}, where $J_i$ is the Bessel function of the first kind of order $i$ and of argument $\delta$. Bessel functions of higher order are omitted for simplicity because they are of decreasing weight and of little relevance for understanding the phase compensation mechanism. After propagation into the fiber spool, the three frequencies experience a relative phase of $\theta$, and the state transforms to $J_{-1}e^{-i\theta}\ket{\nu_{-1}}+J_{0}\ket{\nu_{0}}+J_{1}\ket{\nu_{1}}e^{i\theta}$. After the demultiplexing stage, the control laser passes through a second PM which is driven by an RF signal with an offset phase $\phi$ relative to the RF signal at the input PM. The state $\ket{\Psi}_{\textup{out}}$ after the second PM becomes 
\begin{equation} \label{eq:S1}
\begin{split}
    \ket{\Psi}_{\textup{out}} = & J_{-1}e^{-i\theta}(J_{-1}e^{-i\phi}\ket{\nu_{-2}}+J_0\ket{\nu_{-1}}+J_1e^{i\phi}\ket{\nu_0}) + \\
    {} & + J_0(J_{-1}e^{-i\phi}\ket{\nu_{-1}}+J_0\ket{\nu_{0}}+J_1e^{i\phi}\ket{\nu_1}) + \\
    {} & + J_{1}e^{i\theta}(J_{-1}e^{-i\phi}\ket{\nu_{0}}+J_0\ket{\nu_{1}}+J_1e^{i\phi}\ket{\nu_2}).
\end{split}   
\end{equation}
We used a fiber Bragg grating (FBG) to filter only the component at frequency $\nu_0$, whose intensity $I(\theta,\phi)=|\braket{\Psi_{\textup{out}}|\nu_0}|^2$ is given by 
\begin{equation}
\label{eq:S2}
I(\theta,\phi) = I_0(3-4\cos(\phi-\theta)+2\cos(2(\phi-\theta))),    
\end{equation}
where $I_0=J_0^4$ is the maximum intensity and we used the fact that $J_{i} = (-1)^{-i}J_0$. Equation (\ref{eq:S2}) shows that the interference fringe which is obtained by sweeping the RF phase $\phi$ has an offset $\theta$. The sensitivity with which we can measure $\theta$ is enhanced compared to the use of a standard Mach-Zehnder interferometer due to the term $\cos(2(\phi-\theta))$, and is due to the interference of three frequency modes. In practice, the phase $\theta$ can be used to infer that of Bob's photons because their wavelength ($\lambda_s = 1556.25$ nm) is very close to that of the control laser.
To monitor the phase drift $\theta$ over time, a complete fringe is recorded every $2$ s by sweeping $\phi$, and it is fit with Eq.(\ref{eq:S2}) to extract $\theta$. 
In Fig.\ref{Fig_S1}(c(iii)), we show that $\theta$ monotonically increases during the heating of the fiber spool, which is in agreement with the increase of the optical path length $\Delta L_{\textup{eff}}$ shown in Fig.\ref{Fig_S1}(c(ii)). Indeed, the two quantities are perfectly correlated, as shown in Fig.\ref{Fig_S1}(c(iv)) (Pearson correlation coefficient of $r=0.995$). The correlation is also almost perfect between $\theta$ and $\Delta T$ ($r=0.995$), from which we extract a phase drift per unit length and unit temperature of $285(1)\,\textrm{deg}\cdot\textrm{km}^{-1}{}^{\circ}\textrm{C}^{-1}$, corresponding to $19(1)\,\textrm{deg}\cdot\textrm{km}^{-1}\textrm{GHz}^{-1}{}^{\circ}\textrm{C}^{-1}$.\\
\begin{figure}[t!]
    \centering
    \includegraphics[width = 0.5 \textwidth]{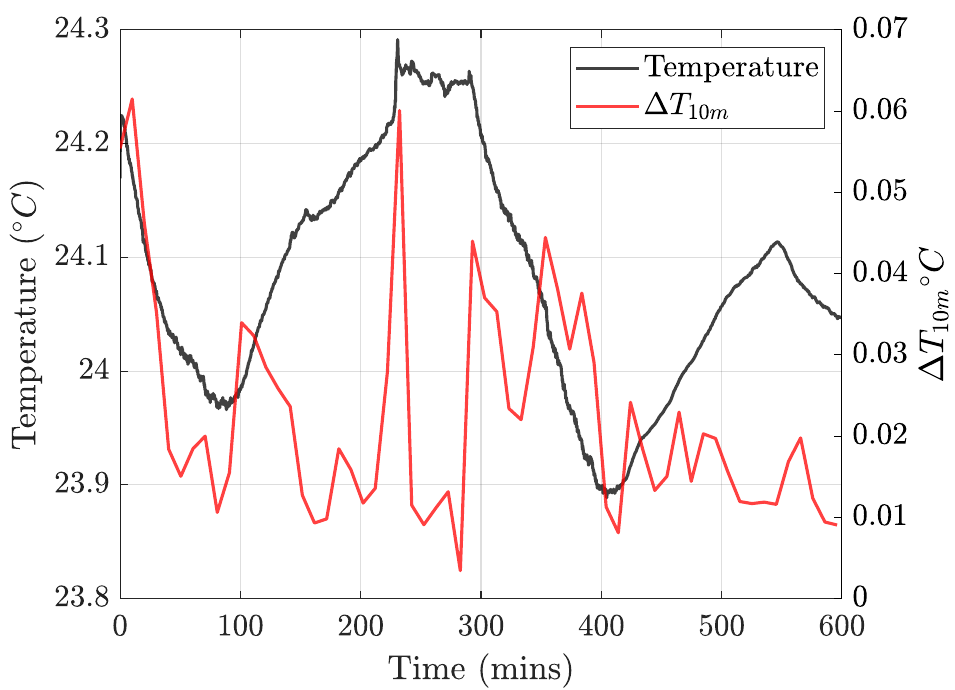}
    \caption{Temperature of the laboratory during a continuous monitoring time of $10$ hours (black). The whole observation time is divided in intervals of $10$ minutes, and in each of them the difference $\Delta T_{10m}$ between the maximum and the minimum value in calculated (red).}
    \label{Fig_S2}
\end{figure}
The experimental results are then compared to those expected by attributing the change in the optical path length to the thermo-optic variation of the effective index and to the thermal expansion of the fiber. The predicted phase drift per unit length and unit temperature is given by $2\pi\Delta\nu(\alpha_{to}+n\alpha_{exp})=211\,\textrm{deg}\cdot\textrm{km}^{-1}{}^{\circ}\textrm{C}^{-1}$, and is very close to the experimental value. The higher slope found in the experiment could be due to the under-estimation of the temperature variation of the fiber, which is sensed in the outermost part of the spool.\\
As described in Section V of the main text, the phase drift can be compensated by rotating the superposition basis by $\theta$. We initially test the active phase compensation system with classical light.  The pulsed laser is replaced by a signal laser tuned at the central wavelength of Bob's photons ($1556.25$ nm). The laser is passed into a PM before entering the fiber spool, which creates three frequency-bins with an offset frequency of $15$ GHz. At the output of the spool, the frequency-bins are mixed by another PM and the signal laser passes through a fiber optic circulator and a FBG to filter detect only the intensity at the base-band frequency. By sweeping the RF phase $\phi_b$, we record interference fringes that depend on the phase offset $\theta_b\sim\theta$ between the frequency-bins at $1556.25$ nm. Figure $\ref{Fig_S1}(d)$ shows examples of the recorded fringes and their time evolution when no offset is added to $\phi_b$ (active phase compensation turned off). It is clear that the fringes do not change in time for a short fiber length ($10$ m), but slowly drifts with a timescale of few minutes when the fiber length is increased to $26$ km. When the (time-varying) phase offset $\theta_b$ is added to $\phi_b$ (active phase compensation turned on), the drift is compensated.\\
The results of this section prove that temperature variations cause a phase drift between the frequency-bins, but the effect was deliberately exaggerated by refrigerating the fiber. Indeed, the transmission tests reported in Section IV of the main manuscript were conducted in a temperature controlled laboratory, where much smaller excursions are expected to occur. These are quantified by continuously monitoring the temperature of the laboratory for $10$ hours, which is reported in Fig.\ref{Fig_S2}. From this record, we evaluated an average variation of $\sim 0.03(1) {}^{\circ} C$ every $600$. Using the linear relation between $\theta$ and temperature change shown in Fig.\ref{Fig_S1}(c(vi)), we extrapolated an average variation $\theta\sim 0.1$ rad per km of fiber every $10$ minutes. These results indicate that the random temperature fluctuations of the environment are sufficient to cause a change of $\theta$ of the order of few radians for fiber spools of tens km of length.
%

\end{document}